# Shedding Light on Moiré Excitons: A First-Principles Perspective

Hongli Guo[1], Xu Zhang[1], and Gang Lu[1*]

[1]*Department of Physics and Astronomy,*
*California State University Northridge,*
*Northridge, California 91330-8268, USA*

**Abstract**

Moiré superlattices in van der Waals (vdW) heterostructures could trap strongly bonded and long-lived interlayer excitons. Assumed to be localized, these moiré excitons could form ordered quantum-dot arrays, paving the way for novel optoelectronic and quantum information applications. Here we perform first-principles simulations to shed light on moiré excitons in twisted $MoS_2/WS_2$ heterostructures. We provide the direct evidence of localized interlayer moiré excitons in vdW heterostructures. The moiré potentials are mapped out based on spatial modulations of energy gaps. Nearly flat valence bands are observed in the heterostructures without "magic angles". The dependence of spatial localization and binding energy of the moiré excitons on the twist angle of the heterostructures is examined. We explore how electric field can be tuned to control the position, polarity, emission energy, and hybridization strength of the moiré excitons. We predict that alternating electric fields could modulate the dipole moments of hybridized moiré excitons and suppress their diffusion in Moiré lattices.

**Introduction**

Van der Waals (vdW) heterostructures formed by vertical stacks of two-dimensional (2D) crystals provide an unprecedented platform to engineer quantum materials with exotic physical properties (unconventional superconductivity(*1, 2*), fractal quantum Hall effect(*3*), Bose–Einstein condensation(*4*)) and novel optoelectronic applications (quantum emitters(*5, 6*), spintronic and excitonic devices(*7-9*)). The most appealing route to engineer a vdW heterostructure is to introduce a lattice mismatch or a rotation misalignment between the 2D layers, resulting in a moiré superlattice with new length and energy scales to be explored for fascinating quantum phenomena(*10-16*). More recently, the formation and controlling of moiré excitons trapped by periodic moiré potentials have attracted significant interest(*17*).

Following initial theoretical predictions(*5, 13*), a number of experimental observations of moiré excitons have been reported in vdW heterostructures of transition metal dichalcogenides (TMDs). (*18-22*) The 2D TMDs feature prominent excitonic effect owing to quantum confinement and reduced dielectric screening(*23-25*). Most TMD heterostructures host long-lived interlayer excitons with electron and hole separated at different layers due to type II band alignment(*26, 27*). When the periodicity of the moiré superlattice exceeds the Bohr radius of the moiré excitons, the excitons could become spatially localized and dispersionless in energy(*12, 28*). The localized moiré excitons are envisioned as programmable solid state analogue of ultracold atoms in optical lattices(*5, 18-20*). And their substantially reduced bandwidths render the moiré lattices an excellent platform for studying exotic quantum phases, from Hubbard model to Mott and Wigner crystals(*29-31*).

Despite the surge of experimental and theoretical research on moiré excitons in vdW heterostructures, first-principles perspectives on this subject are conspicuously scarce, hampered by computational challenges. This is unfortunate because first-principles studies can provide critical insights and a level of atomistic details that are beyond the reach of experiments and phenomenological theories. In particular, first-principles modeling is an indispensable tool in exploring the large and ever-increasing family of vdW heterostructures. In this work, we attempt to fill the gap and carry out first-principles calculations to shed light on moiré excitons in twisted $MoS_2/WS_2$ heterostructures. Using a newly developed computational method, we provide the direct evidence of localized interlayer moiré excitons in vdW heterostructures(*32*). We map out the inter- and intra-layer moiré potentials based on spatial modulations of energy gaps. Nearly flat valence bands are observed in $MoS_2/WS_2$ heterostructures with small twist angles. We examine the dependence of localization and binding energy of the moiré excitons on twist angles of the heterostructures and predict that electric field can be tuned to control the position, polarity, emission energy, and hybridization strength of moiré excitons. Finally, we propose that the formation of hybridized moiré excitons under alternating electric fields could suppress their diffusion in moiré lattices.

**Results**
**Moiré potential and local band gap modulation** Figure 1(a) and (b) show the unit cells of two moiré superlattices, formed by twisting the $MoS_2/WS_2$ bilayer with an angle θ=3.48° and θ=56.52°, respectively. The unit cells have the same lattice constant (5.25 nm) and number of atoms (1626). In both superlattices, there are three local motifs, labeled by A, B and C, that preserve the threefold rotational symmetry ($C_3$). The three motifs play a crucial role in determining the properties of the moiré lattices. The local stacking configurations around these high symmetry points are denoted by $R^{X/Y}$ and $H^{X/Y}$, indicating that X atoms at the $WS_2$ layer sit directly above Y atoms at the $MoS_2$ layer.

A moiré superlattice can be considered as a collection of local stacking motifs – each characterized by a displacement vector ***d***, defined in the primitive unit cell of the untwisted $MoS_2/WS_2$ bilayer (θ=0°) as shown in Figure 2(a). As ***d*** spans the primitive unit cell, all possible stacking motifs in $MoS_2/WS_2$ moiré superlattices can be recovered. In this way, one can map out the maximum amplitude of the moiré potentials, which is the most important property of the moiré superlattices. More specifically, we can calculate the energy band gaps of $MoS_2/WS_2$ bilayer (θ=0°) as a function of ***d***. The band gap $E_g$ is defined as energy difference between the conduction band minimum (CBM) of $MoS_2$ and the valence band maximum (VBM) of $WS_2$. The band gap variation $\delta E_g = E_g - \langle E_g \rangle$ as a function of the stacking displacement ***d*** is shown in Figure 2(b), where $\langle E_g \rangle$ is the average band gap. It is found that the band gap extremes coincide with the high symmetry points of the moiré lattice. As the band gap modulation is often used to characterize the moiré potentials(*13*), we reveal that the maximum value of moiré potentials in $MoS_2/WS_2$ superlattices is ~100 meV at A point whereas the minimum value of the moiré potentials is -160 meV at C point. The moiré potential at B point is slightly shallower than that at C point and the maximum amplitude of the moiré potentials is thus ~260 meV. Note that the amplitude of a specific $MoS_2/WS_2$ structure could be lower than the maximum value of 260 meV.

Similarly, we can also estimate the maximum amplitudes of the intralayer moiré potentials by calculating the intralayer band gap modulations, $\Delta_1$ and $\Delta_2$, defined in Figure 2(a). It is found that the maximum amplitudes of the intralayer moiré potentials are 18 meV for $MoS_2$ and 13 meV for $WS_2$, respectively (Figure S1). As the amplitude of the interlayer moiré potentials is much greater than that of the intralayer moiré potentials, the interlayer moiré excitons are expected to be more localized than the intralayer moiré excitons.

In Figure 2(c), we present the variation of interlayer distance ($\delta h$) as a function of $d$ for MoS$_2$/WS$_2$ superlattices. Here, $\delta h = h - <h>$ with $h$ defined as the vertical distance between the adjacent Mo and W atoms and $<h>$ is its average value. It is found that the maximum $\delta h$ occurs at A point and the minimum at B and C points. It is evident that the modulation of $E_g$ correlates with that of $h$, with the maximum (minimum) interlayer distance corresponding to the maximum (minimum) band gap.

**Flat valence bands and localized moiré excitons** In Figure 3(a), we present the single-particle band structure of the twisted MoS$_2$/WS$_2$ heterostructure with θ=56.52°. A nearly flat VBM band is observed with the bandwidth of 3 meV, and the CBM bandwidth is 25 meV. As comparison, the valence bandwidth of the untwisted MoS$_2$/WS$_2$ heterostructure (θ=0°) is 1500 meV. The corresponding charge densities of the VBM and CBM are shown in Figure 3b. The VBM charge density extends to both layers with the main contribution from $p_z$ and $d_{z^2}$ orbitals, giving rise to significant interlayer mixing. In contrast, the CBM charge density is completely confined to MoS$_2$ layer. Thus, the VBM is modulated by the interlayer moiré potential while the CBM by the intralayer moiré potential. Since the interlayer moiré potential is much deeper than the intralayer potential, the VBM charge density is more localized and the energy band is less dispersive than the CBM. In addition, the in-plane charge densities of the VBM and CBM show that the hole (blue) is more localized than the electron (red). If the single-particle picture were accurate (with negligible excitonic effect), the hole and the electron would have been trapped only at C and B point, respectively.

The similar results are also found in the MoS$_2$/WS$_2$ heterostructure with θ=3.48°, shown in Figure 3(c) and (d). The VBM and CBM bandwidth is 7 and 34 meV, respectively. The hole is trapped at B and C points but extends to both layers. In contrast, the electron is confined at MoS$_2$ layer, but delocalized within the layer. The flat valence bands in both heterostructures result from localized in-plane distributions of the holes, without magic angles(*28*). We have also examined the VBM bandwidths in other MoS$_2$/WS$_2$ heterostructures with different twist angles, summarized in Table S1. In general, as the twist angle increases from 0°, the VBM bandwidth also increases. The band structures and charge densities of VBM and CBM for these heterostructures are shown in Figure S2. Finally, we note that the CBM/VBM charge density distributions in the twisted MoS$_2$/WS$_2$ heterostructures are similar to those in lattice-mismatched MoS$_2$/MoSe$_2$ heterostructure with no in-plane rotation(*33*).

The conventional first-principles approach to capture excitonic effect in semiconductors is GW-Bethe-Salpeter equation (GW-BSE) method(*34-36*) based on many-body perturbation theory. However, the GW-BSE approach is prohibitively expensive for moiré excitons owing to large number of atoms (~1600) in the unit cell. To circumvent the problem, we have recently developed an alternative first-principle method which can provide a reliable description of excitonic effect with much less computational cost(*32, 37-40*). This method is based on time-dependent functional theory(*41, 42*) with optimally tuned and range-separated hybrid exchange-correlation functionals(*43-46*)(details are in Methods session). Using this method, we examine the moiré excitons in twisted MoS$_2$/WS$_2$ heterostructures with different angles. In Figure 4, we present the charge densities of the excitons in the heterostructures with

θ=0° and θ=3.48°. In the absence of a moiré potential (θ=0°), the lowest energy interlayer exciton is delocalized over the heterostructure, with the electron at WS$_2$ and hole at MoS$_2$ layer. In the presence of the moiré potential (θ=3.48°), the low-energy moiré excitons become localized. In particular, the lowest energy moiré exciton ($E_1$=1.885 eV) is trapped at C point which has the lowest moiré potential. The hole distribution (blue) is more localized than the electron (red), consistent with the fact that the VBM is narrower than the CBM. The second lowest energy moiré exciton ($E_2$=1.901 eV) is trapped at B point, again with the hole more localized than the electron. The third lowest energy moiré exciton ($E_3$=1.908 eV) has its electron trapped at A point and hole at B and C points. Evidently, the exciton charge density distributions (Figure 4) deviate from the electron and hole distributions (Figure 3) determined from the single-particle picture, which is a manifestation of the excitonic effect.

Localized moiré excitons are also observed in θ=56.52° MoS$_2$/WS$_2$ heterostructure, shown in Figure 5. The lowest energy moiré exciton has its electron localized at B and hole at C point, respectively. As the energy increases, the electron can also be localized at A and C points, but the hole remains to be trapped primarily at C point.

As the twist angle increases, the moiré potential becomes shallower and the excitons become less localized (Figure S3). To the best of our knowledge, our work provides the first direct evidence of localized moiré excitons in vdW heterostructures from first-principles. Owing to the lateral confinement of moiré potentials, energy levels of the moiré excitons should be quantized. Indeed, we find that for θ=3.48°, the average spacing between the three lowest energy moiré excitons is merely 23 meV, compared to ~200 meV for θ=0° (Figure S4). The small energy spacings imply that the position of the localized moiré excitons may fluctuate at room temperature ($K_B T$ ~ 26 meV).

Finally, we estimate exciton binding energy in various MoS$_2$/WS$_2$ heterostructures, summarized in Table S1. Although the moiré excitons are localized in space, their binding energies are similar to that of delocalized interlayer exciton in the untwisted heterostructure. Thus, we believe that the moiré interlayer excitons could have similar long lifetimes as the delocalized interlayer excitons in the untwisted MoS$_2$/WS$_2$ heterostructure, which has important consequences. Our finding that the exciton binding energy has little dependence on twist angle agrees with the recent work on excitonic properties of local stacking motifs(*47*).

**Electrical tuning of moiré exciton positions** The interlayer exciton features an electric dipole (**P**) that couples with a perpendicular electric field (**ε**) as shown schematically in Figure 6(a). This coupling enables electrical control of exciton properties. On one hand, the electric field could shift the energy levels of MoS$_2$ and WS$_2$ due to the Stark effect. A positive field, pointing from WS$_2$ to MoS$_2$ layer, would increase the energy of MoS$_2$ and lower the energy of WS$_2$, and vice versa for a negative electric field. The band structure changes due to the electric fields are shown in Figure S5 by considering spin-orbit coupling. Therefore, upon a certain positive field, the electron and hole of a moiré exciton could switch layers, forming an interlayer moiré exciton with the opposite polarity. On the other hand, the electric field can also switch the in-plane positions of the moiré excitons. More specifically, because A, B and C points have different interlayer distance (*h*), as shown in Figure 6(b), they would have different coupling energy $E$ according to $E = e\varepsilon h$.

In the absence of electric field, the lowest energy moiré exciton for θ=3.48° is positioned at C point and the second lowest energy moiré exciton at B point with an energy difference $\Delta E = 0.016$ eV (see Figure 4). Since $\Delta h = h_C - h_B = 0.05$ Å, the exciton positions can be switched by a positive field $\varepsilon = \frac{\Delta E}{e \Delta h} = 3.2$ V/nm. This is exactly what the first-principles calculations reveal. As shown in Figure 7(a), the lowest energy moiré exciton shifts from C to B under a positive electric field of 3 V/nm. Interestingly, the electron and hole also switch layers, with the electron now residing at $WS_2$ layer and the hole at $MoS_2$ layer. In other words, the positive field can switch both position and polarity of the moiré excitons. Similarly, as $h_A - h_C = 0.5$ Å, one can apply a negative electric field to switch the exciton positions between A and C. Indeed, our first-principles calculations indicate that a field of -6 V/nm could shift the lowest energy moiré exciton from C to A.

As shown in Figure 6(a), a negative field can reduce the energy gap of the heterostructure, thus the energy of the interlayer excitons. In Figure 7(d), we present the density of excitonic states at three different electric fields. As the electric field varies from -6 to 3 V/nm, the exciton energy can be continuously tuned over a wide range of 600 meV, with red (blue) shift of the exciton energy under the negative (positive) field. To summarize, we have demonstrated that the electric field can be tuned to program the spatial location, polarity and emission energy of moiré excitons, enabling control of quantum information carriers on demand.

**Electrical tuning of exciton hybridization** As shown in Figure 6(a) and Figure S5, under a positive electric field, the CBM/VBM of $MoS_2$ are elevated, and the CBM/VBM of $WS_2$ are lowered. These changes would promote resonant tunneling of the electron and hole across the heterojunction, leading to hybridized excitons(*48, 49*). In Figure 8(a), we present the charge density of the lowest energy moiré exciton in θ=56.5° heterostructure at $\varepsilon$ = 0 and 3 V/nm. While there is a minor hybridization of the hole under zero electric field, the hybridization becomes much more prominent under $\varepsilon$ =3V/nm. In the latter case, the electron (and hole) density spreads equally between the $WS_2$ and $MoS_2$ layers. In other words, the exciton can be considered as both intra- and inter-layer exciton, or alternatively, a hybridized intra- and inter-layer exciton.

In Figure 8(b), we show the charge densities of two lowest energy moiré excitons in θ=3.48° $MoS_2$/$WS_2$ heterostructure under an electric field $\varepsilon$ =3 V/nm. Although the two excitons have the opposite polarity, they have almost the same energies (4 meV difference). Hence, they can be regarded as two approximately degenerate states, and their superposition would have electron and hole spread evenly between the layers. In other words, hybridized moiré excitons can be formed in both heterostructures under the same field. Compared to the interlayer excitons whose dipoles are normal to the plane, the hybridized excitons could orient their dipoles in the plane. In addition, the dipole moment of the hybridized exciton could be reduced from that of the interlayer exciton.

Hybridized moiré excitons have been observed recently in $MoSe_2$/$WS_2$ heterostructures in which the hybridization strength can be tuned continuously with the twist angle(*21, 49, 50*). Our calculations on the other hand show that the hybridization strength can also be tuned by electric field.

**Diffusion of moiré excitons** Although the moiré excitons are localized, they can nonetheless tunnel through the moiré potentials and diffuse over long distances. Thus, the ability to control exciton diffusion - either enhancement or suppression - is of significant interest. In a moiré superlattice, the diffusion of a moiré exciton can be modeled as a series of incoherent hops between the lattice sites with low moiré potentials. The Förster theory is often used to estimate the rates of energy transfer between two weakly coupled dipoles, i.e., moiré excitons at adjacent sites(*51, 52*). In the Förster theory, the relative orientation of the dipole moments can significantly influence the rates of energy transfer. This dependence is captured by the orientation factor κ defined as

$$\kappa = \hat{\mu}_D \cdot \hat{\mu}_A - 3(\hat{\mu}_D \cdot \hat{R})(\hat{\mu}_A \cdot \hat{R}).$$

Here $\hat{\mu}_D$ and $\hat{\mu}_A$ are unit vectors defining the orientation of the dipole moments of the donor and the acceptor excitons and $\hat{R}$ is the unit vector connecting the two. In $MoS_2/WS_2$ heterostructures, $\hat{R}$ always lies in-the-plane, connecting the high symmetry points (A, B and C). As long as the interlayer moiré exciton is not hybridized, $\hat{\mu}_A$ and $\hat{\mu}_D$ are perpendicular to the plane, thus to $\hat{R}$, as shown schematically in Figure 9(a). In this case, the orientation factor κ is 1. When the exciton is hybridized, its dipole moment could lie in-the-plane, as shown in Figure 9(b). In this case, $\hat{\mu}_A$ and $\hat{R}$ would be perpendicular to $\hat{\mu}_D$, yielding $\kappa = 0$. The vanishing κ would suppress exciton diffusion, thus lock the exciton in space. Since the electric field can tune the extent of exciton hybridization, it could also modulate κ value, thus control exciton diffusion.

Finally, we provide a crude estimate for the diffusion of the lowest energy moiré exciton in $MoS_2/WS_2$ heterostructure with θ=3.48°. The lifetime of the exciton is estimated based on the spontaneous emission rate and the exciton hopping rates are determined by the Förster theory (details in the Methods session). Using Monte Carlo simulations, we estimate that the lifetime of the moiré exciton is ~1 ns, similar to the experimental result of 1.8 ns in $MoSe_2/WSe_2$ heterostructure(*26*). On the other hand, localized moiré excitons are expected to have shorter diffusion lengths than the delocalized excitons in the untwisted heterostructure. We find that the diffusion length of the moiré exciton is 0.133 μm, one order of magnitude smaller than that in $WSe_2/MoSe_2$ heterostructure(*8*). The diffusion constant of the moiré exciton is 0.2 $cm^2$/s, which is in line with the experimental results(*8*).

**Discussion**

The large family of 2D materials presents an unprecedented opportunity in engineering quantum materials. With rich physics, unique engineering flexibility and controllability, 2D vdW heterostructures offer a fascinating platform to pursue fundamental science and novel applications. In particular, the presence of strongly bonded, long-lived and localized interlayer moiré excitons in TMD heterostructures could open doors for applications, such as quantum emitters, high performance lasers and twistronics(*53*), etc. Understanding, predicting and ultimately controlling moiré excitons in vdW heterostructures is thus of great scientific importance, but yet highly challenging. In this work, we demonstrate that first-principles simulations are uniquely poised to address the challenges and offer critical insights that cannot be obtained otherwise. For example, we determine the spatial distributions of exciton charge densities from first principles, providing the direct evidence of localized moiré excitons in TMD heterostructures. We map out the interlayer and intralayer moiré potentials that trap the moiré excitons. Nearly flat valence bands are observed without "magic angles". As the twist angle increases from 0°, the moiré superlattices shrink in sizes and the moiré excitons become less localized, but their binding energies remain essentially the same. We show how the electric field can be tuned to control the position, polarity, emission energy, and hybridization strength of the moiré excitons. In particular, we propose that alternating electric fields may lead to oscillating dipoles of the moiré excitons, thus suppress their diffusion.

**Methods**

**First-principles ground state calculations** The ground state properties, including the single-particle band structure, ground state charge density and equilibrium geometry of $MoS_2/WS_2$ heterostructures are modeled by first-principles density functional theory (DFT), implemented in the Vienna *Ab initio* Simulation Package (VASP)(*54, 55*). The Perdew-Burke-Ernzerhof (PBE) exchange-correlation functional(*56*) along with Projector-Augmented Wave potentials(*57, 58*) are used in the self-consistent total energy calculations and geometric optimization. The vdW interaction is considered via the vdW-D2 exchange functional.(*59, 60*) The energy cutoff for the plane-wave basis set is 400 eV. For the band structure calculations of θ=3.48° and θ=56.52° heterostructures, five special *k*-points are sampled along each of the high symmetry lines in the Brillouin zone. The atomic geometry in each moiré superlattice is fully optimized by relaxing all atoms until the residual force on each atom is less than 0.01 eV Å$^{-1}$. A 20 Å vacuum layer is included in the calculations to separate the periodic images of $MoS_2/WS_2$ slab. Spin-orbit coupling (SOC) is taken into consideration in the band structure calculations of the untwisted $MoS_2/WS_2$ heterostructure (θ=0°) under different electric fields (Figure S5). The SOC splitting of the valence bands occurs only at K point, and the splitting at Γ point of VBM is zero. At the same time, the SOC splitting for CBM is negligible, thus the band gap remains indirect, independent of SOC. Thus, we believe that the presence of nearly flat VBM bands in the moiré superlattices with θ=3.48° and θ=56.52° will not be affected if the SOC was considered.

**First-principles excited state calculations** To determine the energies and the many-body wavefunctions of excitons in $MoS_2/WS_2$ heterostructures, we employ a recently developed first-principles approach based on the linear-response time-dependent density functional theory (LR-TDDFT)(*41, 42*) with an optimally tuned, screened and range-separated hybrid exchange-correlation functional (OT-SRSH)(*43-46*). The method has been implemented in conjunction with planewaves and pseudopotentials to study excitonic properties in semiconductors, including graphene fluoride, phosphorene, and 2D perovskites(*32, 37-40*).

The OT-SRSH involves the partition of the Coulombic interaction into a short-range and a long-range contribution based on the following expression(*61*)

$$\frac{1}{r} = \frac{1-[\alpha+\beta erf(\gamma r)]}{r} + \frac{\alpha+\beta erf(\gamma r)}{r}. \quad (1)$$

The range-separated and hybrid exchange-correlation (XC) functional can be expressed as

$$E_{xc}^{RSH} = \alpha E_{xx} + \beta E_{xx}^{LR} + (1-\alpha)E_{KSx} - \beta E_{KSx}^{LR} + E_{KSc}. \quad (2)$$

Where $E_{xx}/E_{xx}^{LR}$ is the Fock-like exact exchange energy, and $E_{KSx}/E_{KSx}^{LR}$ and $E_{KSc}$ are the semilocal Kohn-Sham (KS) exchange and correlation energy, respectively. LR labels the long-range XC terms. $\alpha$ determines the contribution from the exact exchange and $\beta$ controls the contribution from the long-range exchange terms. $\gamma$ is the range-separation parameter. Additionally, $\alpha$ and $\beta$ satisfy the requirement of $\alpha + \beta = \varepsilon_0^{-1}$, where $\varepsilon_0$ is the scalar dielectric constant of the solid, thereby enforcing the correct asymptotic screening of the Coulomb tail(44). The OT-SRSH functional can reproduce the correct long-range electron-electron and electron-hole interactions in solids by choosing the reasonable parameters. In order to reduce the computational cost associated with the Fock-like exchange on large systems, we apply first-order perturbation theory to the range-separated hybrid Kohn-Sham (RSH-KS) Hamiltonian and obtain the first-order corrected RSH-KS eigenvalues and eigenfunctions. (32, 37) We then solve the following non-Hermitian eigenvalue equations of Casida(62),

$$\begin{pmatrix} A & B \\ B^* & A^* \end{pmatrix} \begin{pmatrix} X_I \\ Y_I \end{pmatrix} = \omega_I \begin{pmatrix} 1 & 0 \\ 0 & -1 \end{pmatrix} \begin{pmatrix} X_I \\ Y_I \end{pmatrix} \quad (3)$$

where the pseudo-eigenvalue $\omega_I$ is the $I$-th exciton energy level. The matrix elements of **A** and **B** in the basis of KS states ($ij\sigma$) are given by

$$A_{ij\sigma,kl\tau} = \delta_{i,k}\delta_{j,l}\delta_{\sigma,\tau}(\varepsilon_{j\sigma} - \varepsilon_{i\sigma}) + K_{ij\sigma,kl\tau} \quad (4)$$

$$B_{ij\sigma,kl\tau} = K_{ij\sigma,lk\tau} \quad (5)$$

Here $K$ is the coupling matrix where indices $i$ and $k$ indicate the occupied orbitals, and $j$ and $l$ represent the virtual KS orbitals. According to the assignment ansatz of Casida, the many-body wavefunction of an excited state $I$ can be written as

$$\Phi_I \approx \sum_{ij\sigma} \frac{X_{I,ij\sigma} + Y_{I,ij\sigma}}{\sqrt{\omega_I}} a_{j\sigma} a_{i\sigma} \Phi_0 = \sum_{ij\sigma} Z_{I,ij} a_{j\sigma} a_{i\sigma} \Phi_0 \quad (6)$$

where $z_{I,ij} = (X_{I,ij} + Y_{I,ij})/\sqrt{\omega_I}$; $\hat{a}_{i\sigma}$ is the annihilation operator acting on the $i$th KS orbital with spin $\sigma$ and $\Phi_0$ is the ground state many-body wave function taken to be the single Slater determinate (SD) of the occupied KS orbitals.

In the (TD)DFT-OT-RSH method, there are three parameters, α, β, and γ, needed to be specified. α controls the short-range exact exchange and we choose α = 0.12 in this work. β is chosen to satisfy the requirement α + β = 1/$\varepsilon_0$. The scalar dielectric constant of MoS$_2$/WS$_2$ heterostructure $\varepsilon_0$ is set to 1.0, which is the correct asymptotic limit for screening in 2D materials(63). The other two independent parameters α = 0.06 and γ = 0.03 are determined by reproducing the fundamental gap of MoS$_2$/WS$_2$ heterojunction obtained from the one-shot GW calculations. The exciton binding energy, which is the difference between the fundamental gap and optical gap $E_b = E_g - E_{opt}$ is 0.50 eV, in a good agreement with previous experimental and theoretical results(64-66). Thanks to the large unit cells (1626 atoms) in our LR-TDDFT calculations of the moiré excitons, only Γ point is sampled in the Brillouin zone. We do not consider SOC correction in the LR-TDDFT calculations for the following reasons: (1) The SOC effect is known to be small at Γ point with zero Rashba splitting for parabolic bands(*67*). (2) The SOC has a negligible effect on the exciton charge densities. (3) TDDFT-OT-RSH calculations with SOC for such a large system (1626 atoms) is beyond our computational capability.

**Exciton lifetime and diffusion constant calculation** We model the exciton diffusion as random walks using Monte Carlo (MC) simulations(*68*). For a moiré exciton $I$, one can generate an event table with $M + 2$ transition probabilities ($M$ is the number of the nearest-neighbor moiré excitons): transition probabilities from the exciton $I$ to $M$ neighboring excitons, $P_{1,2,\cdots,M} = k_{IJ} \times \Delta t$, with $J = 1,2,\cdots,M$; annihilation probability, $P_{M+1} = k_{I0} \times \Delta t$; and the probability to remain at the same state $I$, $P_{M+2} = 1 - (P_1 + P_2 + \cdots + P_{M+1})$. $\Delta t$ is the time step of the MC simulations, which is set to be 10 fs. Here, we use Förster resonance energy transfer (FRET)(*52, 69*) to determine the exciton transition rate $k_{IJ}$ as

$$k_{IJ} = \frac{1}{\hbar^2} \frac{\langle \kappa^2 \rangle}{R^6} \mu_I^2 \mu_J^2 \delta(\Delta\omega) \quad (7)$$

where $\mu_I$ and $\mu_J$ are the transition dipole moments of the states $I$ and $J$, respectively. κ is a geometric factor depending on the relative orientation of the dipoles. R is the distance between the center of mass of the two excitons and Δω is the energy difference between the two excitons. The exciton annihilation lifetime is estimated based on the spontaneous emission rate as

$$k_{I0} = \frac{4n(\omega_i - \omega_j)^3 |\mu_I|^2}{3c^3} \quad (8)$$

where $c$ is the vacuum speed of light and $n$ is the refractive index which takes a value of 1.5 for $MoS_2/WS_2$. With the event table, a diffusion trajectory of the moiré exciton $I$ is obtained by executing MC moves until the exciton is annihilated. From each trajectory, one can determine the lifetime $t$ (the number of MC moves multiplied by $\Delta t$) and the maximum distance $d$ of exciton diffusion. Averaging over all trajectories with the same initial exciton position gives the exciton diffusion length, lifetime, and diffusivity by $L_D = \langle d \rangle$, $\tau = \langle t \rangle$ and $D = \frac{\langle d^2 \rangle}{4\tau}$, respectively, where the brackets indicate the average.

**Acknowledgements**

The work was supported by National Science Foundation (DMR1828019) and Army Research Office.

**Author contributions**

G.L. designed the study and H.G. carried out calculations with the assistance of X.Z. G.L. supervised the research and wrote the manuscript with H.G. All authors discussed the results.

**Competing interests**

The authors declare that they have no competing interests.

**Data and materials availability**

All data needed to evaluate the conclusions in the paper are present in the paper and/or the Supplementary Materials. Additional data related to this paper may be requested from the authors.

**Figures**

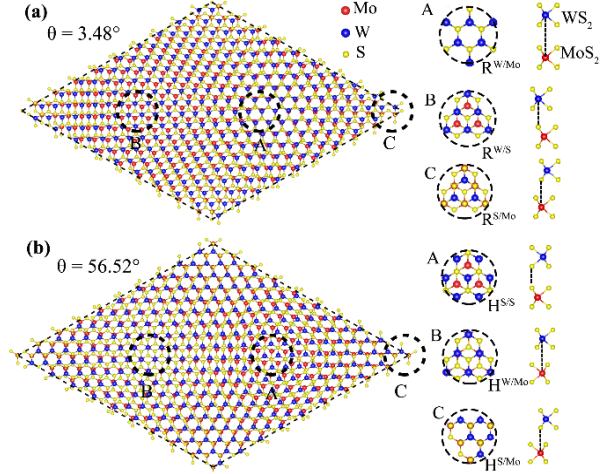

Figure 1: The unit cell of the moiré superlattice formed by a twisted $MoS_2/WS_2$ heterostructure with angle $\theta=3.48°$(a) and $\theta=56.52°$(b). The stacking configurations of the three local motifs, A, B, and C, are shown on the right.

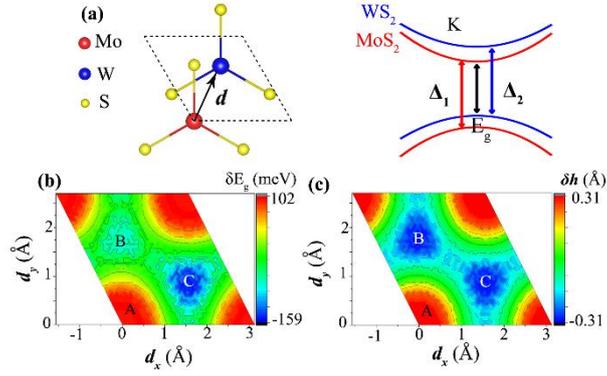

Figure 2: (a) The definition of in-plane displacement $\boldsymbol{d}$ in the primitive unit cell of the $MoS_2/WS_2$ heterostructure with $\theta=0°$ (left). The definition of three band gaps, $E_g$, $\Delta_1$ and $\Delta_2$ (right). (b) Variation of $MoS_2/WS_2$ band gap ($E_g$) as a function of $\boldsymbol{d}$, showing three extremes at A, B and C points. (c) Variation of the interlayer distance, $\delta h$ as a function of $\boldsymbol{d}$, with the extremes at A, B and C points.

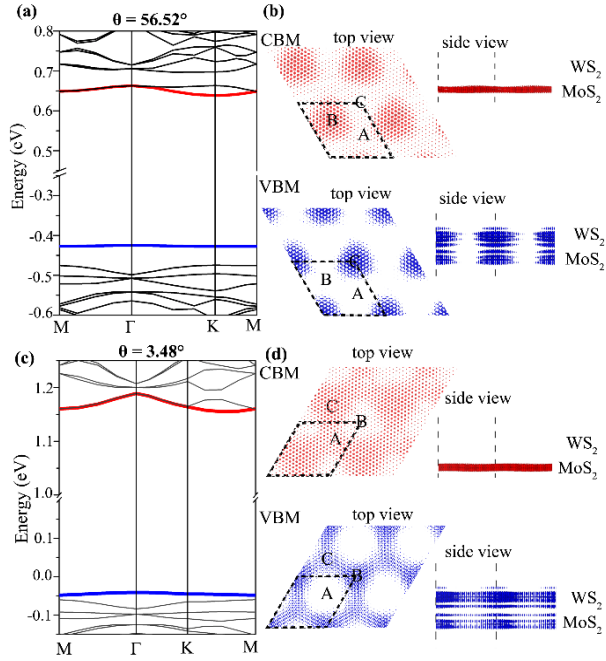

Figure 3: (a) The single-particle band structure for the MoS$_2$/WS$_2$ heterostructure with θ=56.52°. The CBM and VBM bands are shown in red and blue, respectively. (b) The top and side view of the charge density of the CBM and VBM bands for the heterostructure. The unit cell of the moiré lattice is indicated by the dashed box. (c) The band structure for the MoS$_2$/WS$_2$ heterostructure with θ=3.48°. (d) The top and side view of the charge density of the CBM and VBM bands for the heterostructure.

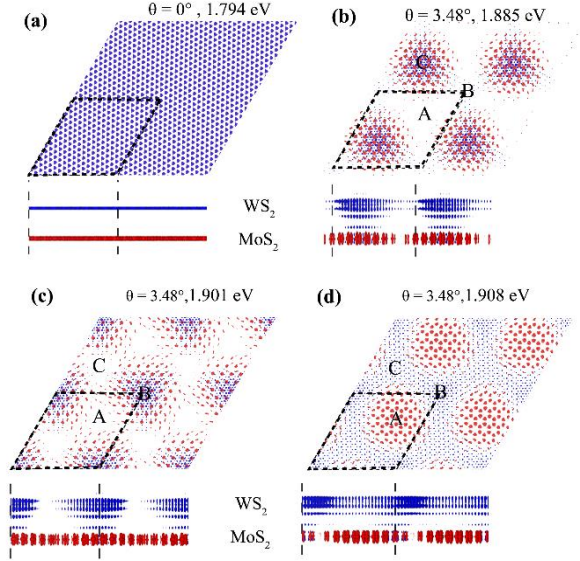

Figure 4: (a) The charge density and energy for the lowest energy exciton in the MoS$_2$/WS$_2$ heterostructure with θ=0° (upper panel: top view; bottom panel: side view). (b), (c), (d) The charge density and energy for the three lowest energy moiré excitons in the twisted MoS$_2$/WS$_2$ heterostructure with θ=3.48° (upper panel: top view; bottom panel: side view). The dashed box indicates the unit cell of the moiré superlattice. Red and blue color represents the charge density of the electron and the hole, respectively. All iso-surface value is set at 0.0001 e/A$^3$.

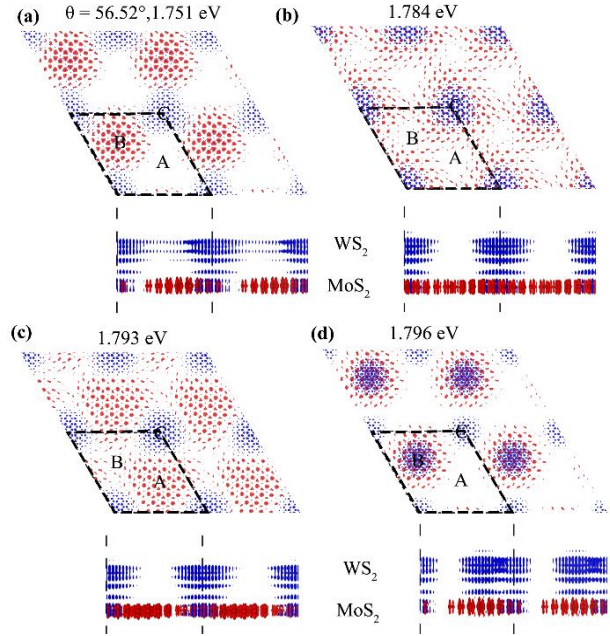

Figure 5: (a), (b), (c), (d) The charge density and energy for the four lowest energy moiré excitons in the twisted MoS$_2$/WS$_2$ heterostructure with θ=56.52° (upper panel: top view; bottom panel: side view). The dashed box indicates the unit cell of the moiré superlattice. Red and blue color represents the charge density of the electron and the hole, respectively. All iso-surface value is set at 0.0001 e/A$^3$.

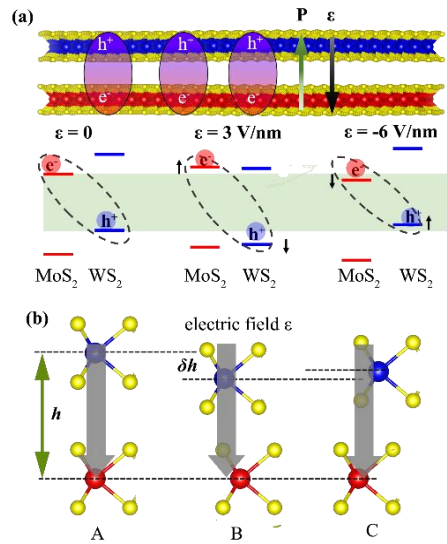

Figure 6. (a) (Up) The schematic picture of MoS$_2$/WS$_2$ heterostructure under a perpendicular electric field ε. The dipole moment of the interlayer exciton is indicated by **P**. (Bottom) The electric tuning of the type II band alignment of the heterostructure. (b) The variation of the interlayer distance $h$ at the A, B and C points for MoS$_2$/WS$_2$ heterostructure with θ=3.48°.

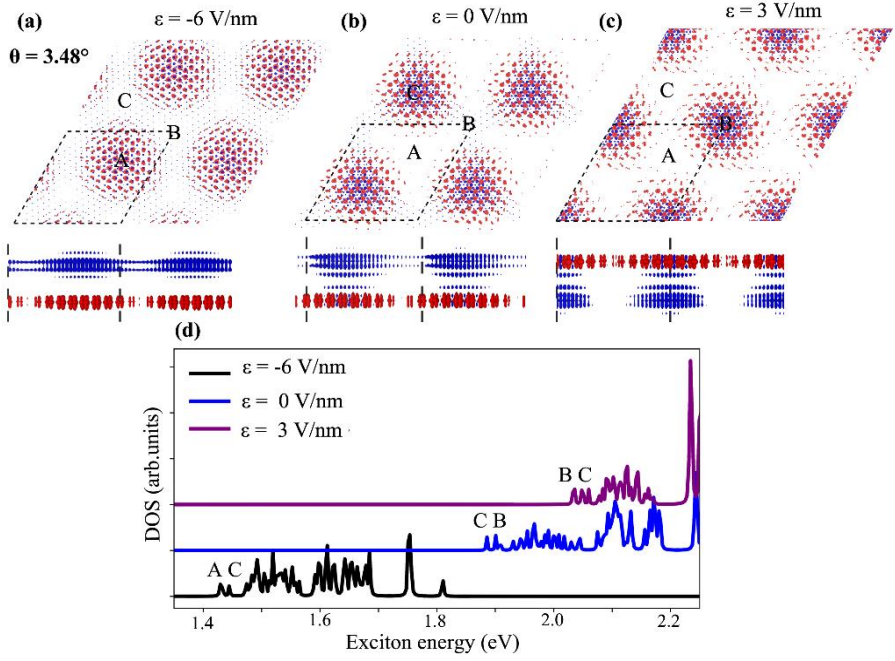

Figure 7: (a), (b), (c) The charge density distribution (top and side view) for the lowest energy moiré exciton in MoS$_2$/WS$_2$ heterostructure with θ=3.48° under different electric fields. Red and blue color represents the charge density of the electron and the hole, respectively. (d) The density of states (DOS) for the excitons under different electric fields, showing field-tunable exciton transition energies. The spatial locations (A, B, and C) for the two lowest energy excitons are indicated.

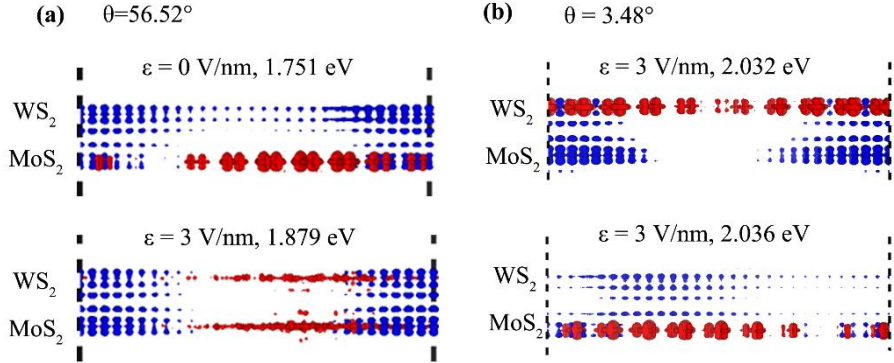

Figure 8: (a) The side view of the charge density distribution for the lowest energy exciton in MoS$_2$/WS$_2$ heterostructure with θ=56.5° under different electric fields. Red and blue color represents the charge density of the electron and the hole, respectively. (b) The side view of the charge density distribution for the two lowest energy excitons in MoS$_2$/WS$_2$ heterostructure with θ=3.48° under the electric field of ε =3 V/nm.

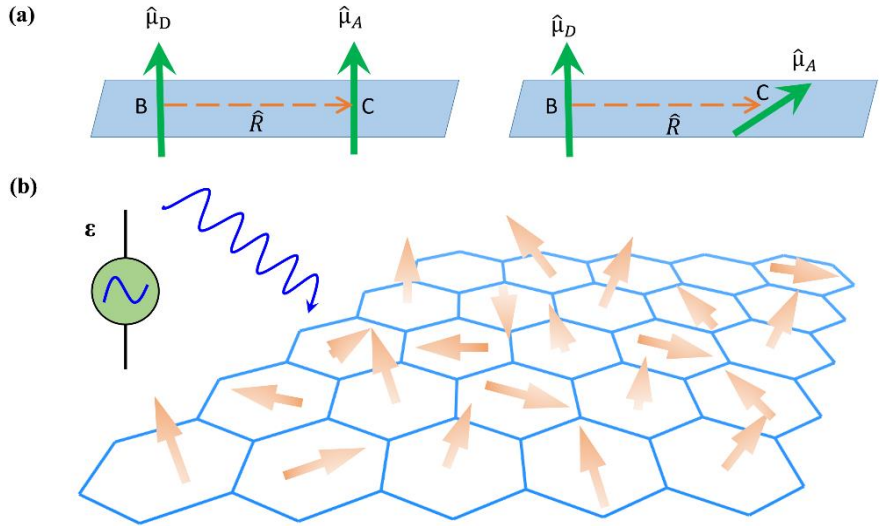

Figure 9: (a) Schematic diagram showing the dipole directions of a diffusing moiré exciton from B to C points in a moiré superlattice. (b) Schematic picture depicting the fluctuation of the dipole moment of a diffusing moiré exciton under an alternating electric field.

# Supplementary Materials

# Shedding Light on Moiré Excitons: A First-Principles Perspective

Hongli Guo[1], Xu Zhang[1], and Gang Lu[1*]
[1]Department of Physics and Astronomy,
California State University Northridge,
Northridge, California 91330-8268, USA

Table S1. The variation of interlayer distance ($\delta h$), VBM bandwidth, and exciton binding energy as a function of twist angle.

| Angle (°) | 3.5 | 6 | 9.56 | 21.7 | 32.2 | 0 | 56.5 |
|---|---|---|---|---|---|---|---|
| $\delta h$ (Å) | 0.54 | 0.44 | 0.22 | 0.02 | 0.02 | 0.00 | 0.58 |
| VBM bandwidth (meV) | 7 | 25 | 111 | 617 | 345 | 1500 | 3 |
| Exciton binding energy (eV) | 0.43 | 0.42 | N/A | 0.43 | 0.39 | 0.50 | 0.42 |

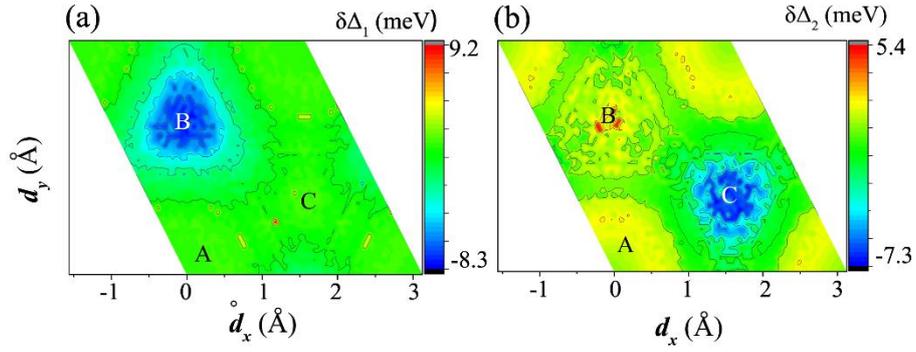

Figure S1: Variation of intralayer band gaps $\Delta_1$ (a) and $\Delta_2$ (b) as a function of $d$.

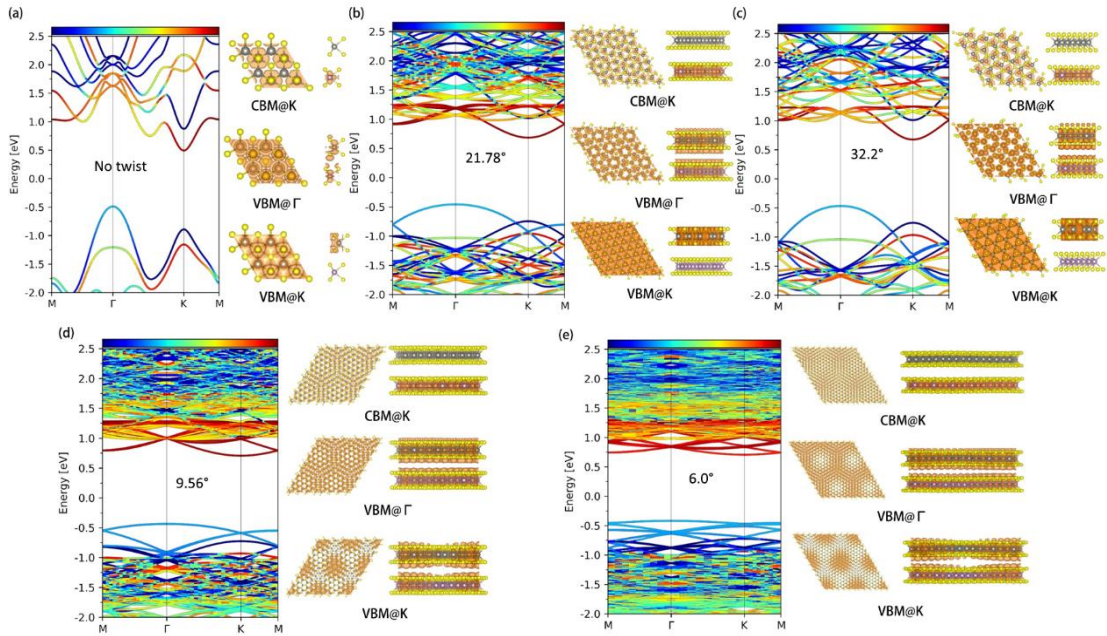

Figure S2: Band structures and charge densities of VBM and CBM for $MoS_2/WS_2$ heterostructures with different twist angles.

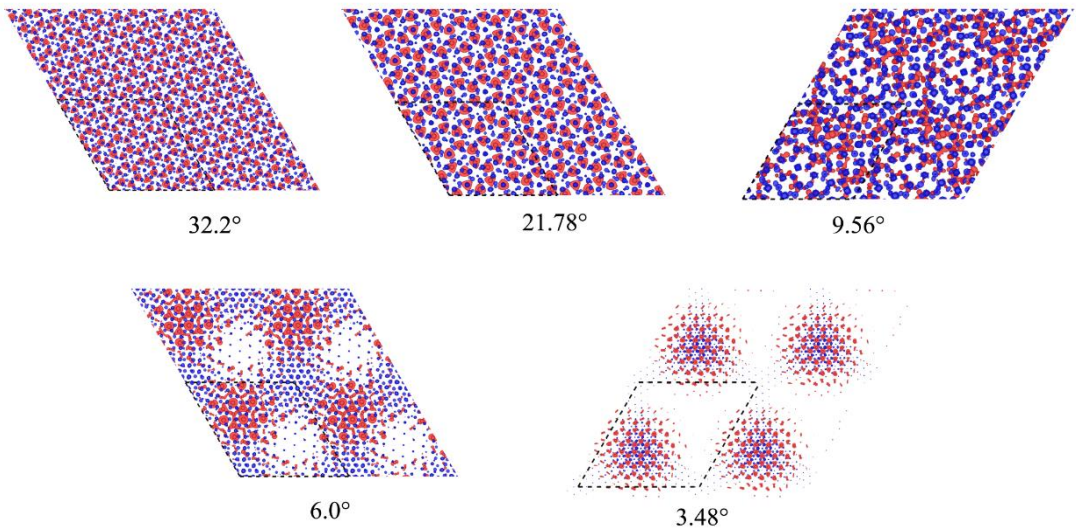

Figure S3: Exciton charge densities of $MoS_2/WS_2$ heterostructures with different twist angles.

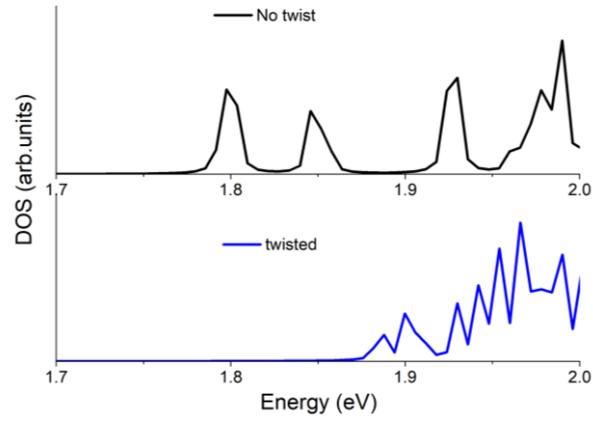

Figure S4: Excitonic density of states (DOS) for $MoS_2/WS_2$ heterostructures with θ=0° (no twist) and θ=3.48°. The DOS of twisted heterostructure shows denser energy spacing.

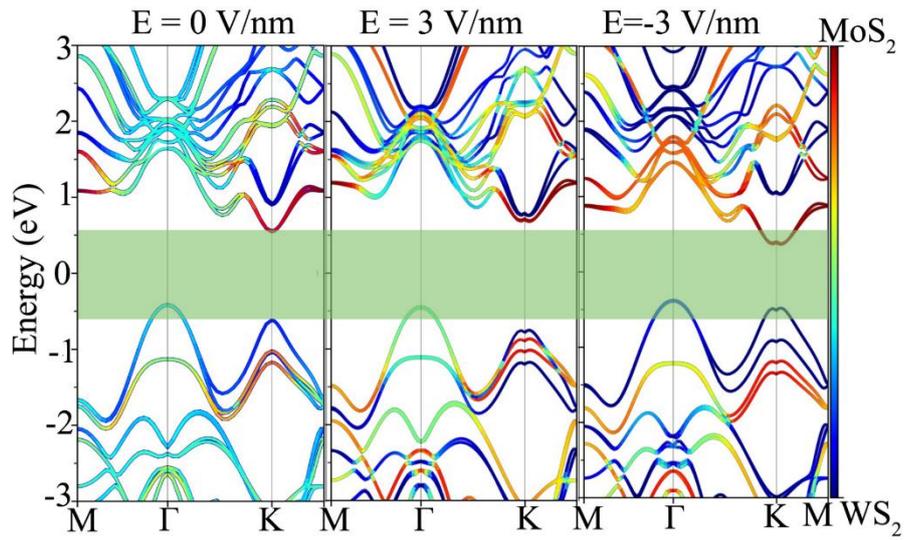

Figure S5: PBE+SOC band structures of $MoS_2/WS_2$ heterostructures with θ=0° under different electric fields, showing the energy level shifts, that are consistent with Figure 6(a).